\documentclass[aps,prb,twocolumn,amsmath,amssymb,floatfix,groupedaddres]{revtex4-1}
\usepackage{graphicx}
\usepackage{float}
\usepackage{dcolumn}
\usepackage{bm}
\usepackage{bbold}
\usepackage{stmaryrd}
\usepackage{latexsym}
\usepackage{amssymb}
\usepackage{amsfonts}
\usepackage{amsmath}
\usepackage{mathtools}
\usepackage{fancybox}
\usepackage{color}
\usepackage{epsfig}
\usepackage[breaklinks=true,colorlinks,citecolor=blue,linkcolor=blue,urlcolor=blue]{hyperref}
\usepackage[normalem]{ulem}

\begin{document}
\title{Majorana zero modes in a quantum wire platform without Rashba spin-orbit coupling}
\author{\"Omer M. Aksoy}
\email{aksoyo@ethz.ch}
\affiliation{Institute for Theoretical Physics, ETH Zurich, 8037 Zurich, Switzerland}
\author{John R. Tolsma}
\affiliation{Institute for Theoretical Physics, ETH Zurich, 8037 Zurich, Switzerland}
\date{\today}
\begin{abstract}
We propose a platform for engineering helical fermions in a hybridized double-quantum-wire setup. When our setup is proximity coupled to an $s$-wave superconductor it can become a class $D$ topological superconductor exhibiting Majorana zero modes. The goal of this proposal is to expand the group of available Hamiltonians to those without strong Rashba spin-orbit interactions which are essential to many other approaches. Furthermore, we show that there exist electron-electron interactions that stabilize fractional excitations which obey $\mathbb{Z}^{\,}_{3}$ parafermionic algebra. 
\end{abstract}
\maketitle

\section{Introduction}
\label{sect:Intro}
Following two insightful theoretical proposals to realize Majorana zero modes (MZMs) in quantum wires with strong Rashba spin-orbit coupling~\cite{lutchyn2010majorana,oreg2010helical, lutchyn2011search}, several experiments have observed zero-bias conductance peaks which provide strong support their existence~\cite{mourik2012signatures,das2012zero}. Despite the success of this particular proposal, and a large number of subsequent works which propose alternative platforms which might support MZMs~\cite{alicea2012new,lado2018two,klinovaja2013topological,zeng2018quantum,lei2018ultrathin}, it remains an open question as to which setup will allow experiments to demonstrate the ultimate ``smoking-gun'' signature of MZMs: braiding of MZMs to reveal their non-Abelian statistics. One of the drawbacks of the Rashba quantum-wire approach is the variable strength of the Rashba coupling amongst materials and the fact that Rashba spin-orbit coupling terms in the Hamiltonians break inversion symmetry.

In this paper, we propose a scheme for engineering MZMs in a quantum-wire platform which does not rely on the appearance of a Rashba spin-orbit coupling term in the low-energy $\vec{k} \cdot \vec{p}$ Hamiltonian. 
Instead, we make use of the large variability of effective Land\'e $g$ factors in real materials (including negative values) and the orbital-coupling of electrons to external magnetic fields. The effective Land\'e $g$ factor of electrons in Bloch bands of periodic solids has for a long time~\cite{luttinger1956quantum,cohen1960g} been known to deviate from the bare value of $g_{\rm bare} \sim 2$. The interplay of lattice strain, quantum confinement, electron interactions, and atomic spin-orbit coupling can lead to Land\'e $g$ factors nearly two orders of magnitude away from the bare $g$-factor value, and does not require inversion symmetry being broken~\cite{winkler2017orbital}. A recently proposed method to calculate the effective Land\'e $g$ factors using first-principles codes~~\cite{dai2017negative,song2015first} will help to further exploit this highly-tunable degree of freedom in engineering nanostructure proposals for MZMs.

\begin{figure}[h!]
\includegraphics[width=0.95\linewidth]{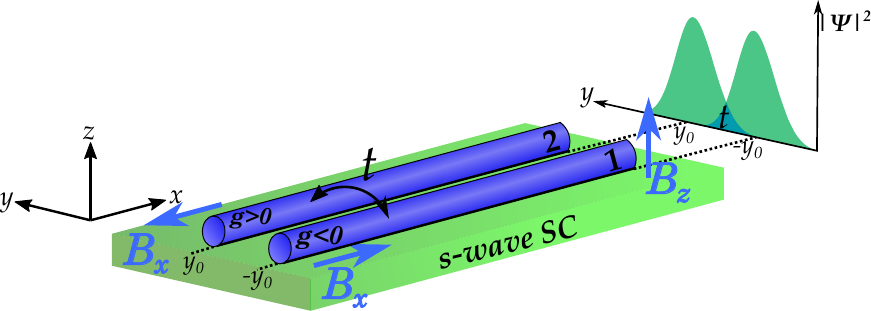}
\caption{(Color online) Schematic diagram of the double nanowire platform. The wires have effective Land\'e $g$ factors of opposite signs. The wires are closely spaced along the $y$ direction by a distance $2 y_0$ such that a small interwire electron tunneling-amplitude, $t$, is present. The presence of a strong magnetic field in the $z$ direction, and the particular gauge choice, $\vec{A} = (-yB_z,zB_x\rm{sgn}(y),0)$, leads to shifts in the band-energy minima which are in opposite directions in momentum space for each of the two wires [see e.g. Fig.~\ref{fig:Two} (a)]. These \emph{orbital shifts} imitate the Rashba spin-orbit coupling effect. The cooperative effect of a small staggered magnetic field in the $x$ direction, $B_x$, and the small spin-conserving interwire tunneling amplitude, $t$, opens a gap between the lowest-energy and second-lowest-energy subbands at $p_x=0$, and gives a low-energy effective Hamiltonian similar to the edge-states of two-dimension quantum spin Hall insulators. When both wires are proximity-coupled to an $s$-wave superconductor, Majorana quasiparticles can emerge. When strong electron-electron interactions are accounted for, fractional excitations become possible. \label{fig:One}}
\end{figure}

Our paper is organized as follows. In Sec.~\ref{Sect:One} we introduce a model of two tunnel-coupled quantum wires with opposite sign effective Land\'e $g$ factors. We explore the energy spectrum of this setup in an external magnetic field while accounting for both Zeeman-coupling and the orbital-coupling of the electrons to the electromagnetic gauge field. We find that the combined effect of spin-conserving interwire hopping and a small magnetic field directed along the wires can lead to emergent helical fermions, and we derive a low-energy Hamiltonian to describe them. In Sec.~\ref{Sect:Two} we consider the effects of interwire and intrawire pairing potentials, which we take to originate from proximity coupling to the bulk $s$-wave superconductor. We analytically construct the wave functions of the MZMs which appear at the two ends of the tunnel-coupled wires. We then remove the restriction of small external magnetic fields along the wires and calculate the topological phase diagram of our model using a Pfaffian representation for the invariant. In Sec.~\ref{Sect:Three} we analyze the effect of strong electron-electron interactions by using bosonization techniques and identify the conditions for fractional excitations to appear. Finally, in Sec.~\ref{Sect:Four}, we conclude and discuss possible realizations.	
\section{Engineering Helical fermions with Magnetic Fields}
\label{Sect:One}
In this section we describe how to engineer helical fermions in a one-dimensional system which does not rely on either Rashba spin-orbit coupling\cite{oreg2010helical,oreg2014fractional,lutchyn2011search} or being on the edge of a topological insulator phase of higher dimension \cite{fu2009josephson,hsu2018majorana}. We demonstrate that a low-energy subspace of helical fermions can emerge in platforms consisting of two quantum wires with opposite sign effective Land\'e $g$ factors when the wires are under the combined effects of orbital coupling to an external magnetic field, and finite spin-flip interwire electronic hopping. After the identification of the low-energy Hamiltonian governing the helical fermions, we add proximity coupling to an $s$-wave superconductor and discuss the topological phase diagram of the system. 

We begin by considering two quantum wires which are sufficiently narrow that only a single (spin-degenerate) subband is occupied in each wire. The two wires extend along the $x$-direction and are spatially separated at $-y_0$ and $y_0$ in the $y$-direction (see Fig.~\ref{fig:One}) and experience finite transverse and staggered in-plane magnetic fields, $B_z$ and $B_x$, respectively. We label the former as wire 1 and the latter as wire 2. Before we consider adding interwire hopping, strong electron-electron interactions, or proximity coupling to superconductors, each wire's electronic properties are determined by their effective mass, $m_{1(2)}$, and their effective Land\'e factor, $g^{\mathrm{eff}}_{1(2)}$. The single-particle Hamiltonian of an electron in wire $j$ is (at this point) simply given by
\begin{equation}
{\rm H}_{{\rm wire}\, j} = \frac{1}{2 m_j}\left[p_x-\frac{e}{c}A_x(\vec{r}_j)\right]^2 - \frac{g^{\mathrm{eff}} _{j}\mu_B}{\hbar}\vec{S}\cdot\vec{B}~, \label{eq:Landau}
\end{equation}
where $A_x (\vec{r}_j)$ is the $x$ component of the vector potential at the position of an electron in the $j$th wire. The first term in Eq.~(\ref{eq:Landau}) is the kinetic energy. It is shifted in momentum due to the orbital coupling to the transverse magnetic field. The second term is the Zeeman coupling, where $\mu_B = e \hbar/(2 m_ec)$ is the Bohr magneton with bare electron mass $m_e$. For simplicity we set effective mass in both wires to be equal, $m_i = m_j = m$. 
\begin{figure}[h]
\includegraphics[width=0.95\linewidth]{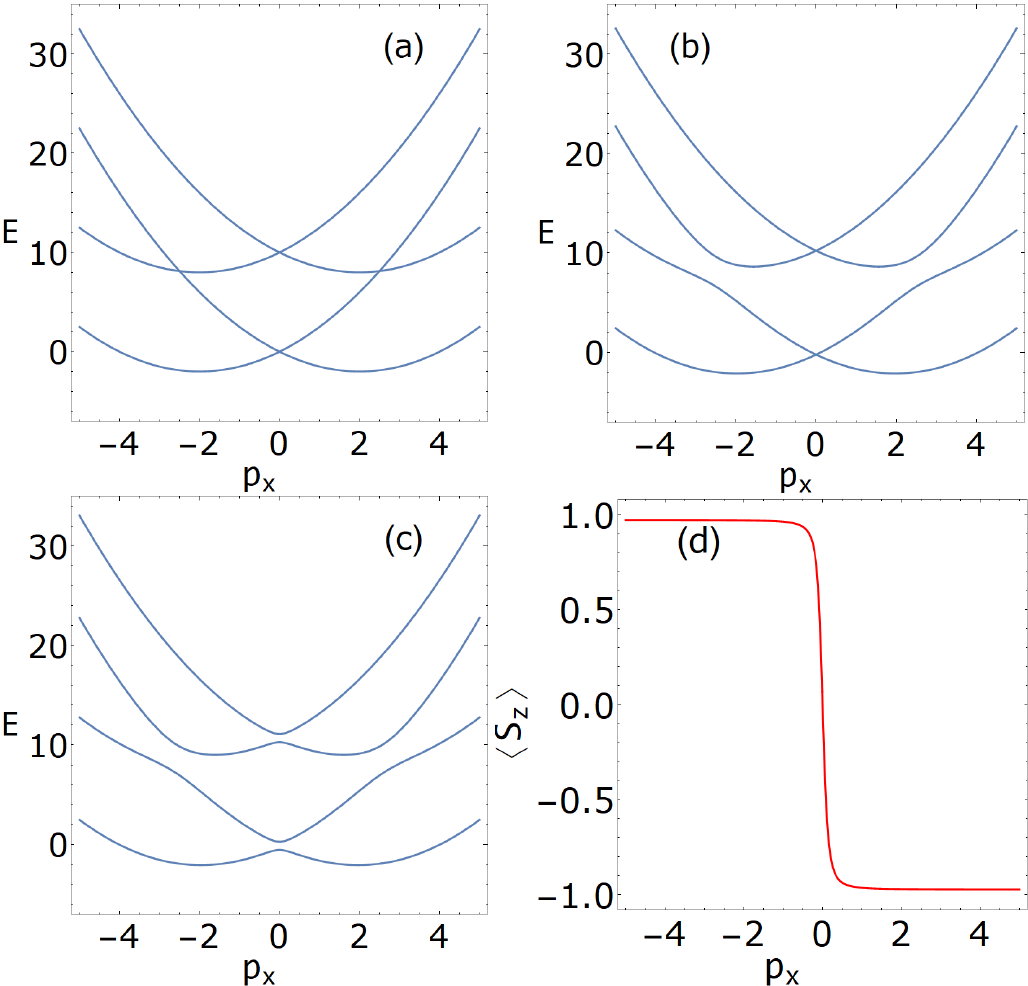}
\caption{(Color online) Figures describing how the two-wire model yields \emph{helical} fermions prior to introducing a finite superconducting proximity coupling. Panels (a)--(c) are three schematic figures which exhibit how the energy versus momentum band structure of the two-wire setup evolves as $B_z$, $t$, and $B_x$ are given non-zero values, respectively. For definitions of $B_z$, $t$, $B_x$, and their values, refer to the main text. In panel (d), we demonstrate that the lowest-energy band describes \emph{helical} fermions by plotting the projection of the electron spin in the $z$ direction against momentum. Transverse magnetic field, effective electron mass, and orbital shift are set to $B_z = 5$, $m=1$ and $A_x = 2$.  (a) Spectrum of the uncoupled bands, i.e., $t=B_x=0$. Position of the band minima are shifted due to the orbital shift $A_x$ and lowest energy value for each band are shifted due to Zeeman coupling. (b) Energy spectrum when interwire hopping is turned on, $t=1.5$ and $B_x = 0$, which results in a gap opening, separating four bands into two lower and higher energy bands. (c) Energy spectrum when in-plane magnetic field is turned on, $B_x = t = 1.5$. This result in gaps opening at $p_x=0$ and isolating the lowest lying band. (d) Expectation value of the $z$ component of spin in units of $\frac{\hbar}{2} $ as a function of momentum in the lowest lying band. States with positive(negative) momenta are spin down (up), forming a helical liquid. \label{fig:Two}}
\end{figure}

The first important ingredient in our model is that we retain the (often overlooked) orbital-coupling of the electron to the magnetic field (see Ref.\ \cite{lutchyn2018topological} for an example where orbital effects are taken into consideration). We make the gauge choice $\vec{A} = (-yB_z,zB_x\rm{sgn}(y),0)$. Because of this orbital-coupling, the minima of the spectrum of ${\rm H}_{{\rm wire}\, j}$ is shifted towards positive momenta for wire $j=1$ and towards negative momenta for wire $j=2$. The second important ingredient in our platform is to use two wires which have opposite sign Land\'e factors, ${\rm sgn}(g^{\mathrm{eff}}_{1}) = -{\rm sgn }(g^{\mathrm{eff}}_{2})$. The transverse magnetic field in the $z$ direction results in Zeeman splitting where the spin-down band in wire 1 is lowered in energy and the spin-up band in wire 2 is lowered in energy. These two bands, which have opposite spin, will form the basis for our low-energy subspace supporting helical fermions. At this point, our two wires are uncoupled (i.e., unhybridized), and furthermore, the degeneracy between these two bands at $p_x =0$ needs to be lifted before we can obtain the desired helical fermions. We can accomplish both of these goals by taking the wires to be spaced closely enough for there to be a finite overlap between the electron wave functions, which yields an interwire \emph{spin-conserving} hopping amplitude $t$. The gap opening at $E \sim 10$ near $p_x = \pm {mB_z}/{A_x}$ in Fig.~\ref{fig:Two}(b) demonstrates that this hopping amplitude acts to hybridize electronic wave functions of the two wires which have the same spin, but does not hybridize electrons of opposite spin, and therefore does not open a gap at the $p_x = 0$ degeneracy point between the lowest energy bands which are of opposite spin. To allow spin-flip transitions when electrons hop between wires, we add a finite $B_x$ field, which is staggered in space to account for opposite Land\'e factors of two wires. Such a staggered magnetic field can be realized by nanomagnets \cite{egger2012emerging,karmakar2011controlled}. After this addition, we observe a finite gap opening between the lowest two energy bands in Fig.~\ref{fig:Two}(c) for $p_x = 0$. When both $t$ and $B_x$ are finite, the resulting second-quantized Hamiltonian is 
\begin{equation}
{\rm H}_{\Delta = 0} = \int \psi^\dagger  \left[\mathcal{H}_{\Delta = 0}\right]  \psi \, {\rm dx} ~,
\end{equation}
where $\psi=(\psi_{1\uparrow},\psi_{1\downarrow},\psi_{2\uparrow},\psi_{2\downarrow})^T$ and
\begin{equation}\label{eq:HamNOdelta}
\mathcal{H}_{\Delta = 0} = \frac{p_x^2+A_x^2}{2m} - \mu-\frac{A_xp_x}{m}\rho_z + B_z\sigma_z\rho_z + B_x\sigma_x + t\rho_x~.
\end{equation}
The Pauli matrices $\sigma$ and $\rho$ act on the spin and wire degrees of freedom, respectively, and we have emphasized that no proximity induced pairing potential has yet been included (i.e., ${\Delta = 0}$). In Eq.~(\ref{eq:HamNOdelta}) and Fig.~\ref{fig:Two}, the units are chosen such that $e=c=\mu_B=\hbar=1$ and the Land\'e factors are for simplicity chosen to be equal in magnitude, and can thus be absorbed into the components of the vector $\vec{B}$. When we compare Eq.~(\ref{eq:HamNOdelta}) with the Hamiltonians of quantum wires with Rashba spin-orbit coupling~\cite{oreg2010helical,oreg2014fractional,klinovaja2014time}, we observe that the term $(A_x/m) p_x \rho_z$ acts like spin-orbit coupling because of the simultaneous presence of the large $B_z\sigma_z\rho_z$ term. Specifically, the latter term acts to lower the energy of the up-spin (down-spin) bands in wire 2 (wire 1), such that states with opposite spin have equal energy but their wave function support is now primarily residing in different wires. When only the lowest-lying band is occupied in each wire, this $B_z\sigma_z\rho_z$ Zeeman term enables the $\rho_z$ matrix to effectively act on the spin-channel. Note that Hamiltonian ${\rm H}_{{\rm wire}\, j}$ in Eq. \eqref{eq:Landau} is symmetric under inversion with respect to the center of the $j$th wire.
Therefore, we do not rely on inversion symmetry breaking (on neither bulk nor structural inversion asymmetries) within each wire to produce a sizable spin-orbit coupling
\footnote{We acknowledge that atomic spin-orbit coupling is still required to 
	generate Land\'e $g$ factors with opposite signs.}. However, the overall system is still not 
inversion symmetric due to the staggered magnetic field.

In the lowest energy band, the expectation value of the spin $z$ component, $\langle S_z \rangle$, takes values with $\langle S_z \rangle = {\rm sgn}(p_x)$, i.e., momentum states of positive(negative) sign have definite down(up) spin states, as shown in Fig. \ref{fig:Two} (d). This illustrates the formation of  $helical$ fermions in the isolated lowest energy band. The Fermi energy sits inside the gap between the lowest energy and second-lowest energy bands in Fig.~\ref{fig:Two} when the chemical potential is tuned via electrostatic gates near the value
\begin{equation}
\mu_{\rm gap} = \frac{A_x^2}{2} - \frac{\sqrt{B_z^2 + (B_x+t)^2}+\sqrt{B_z^2 + (B_x-t)^2}}{2} ~.\label{eq:mu}
\end{equation} 

\begin{figure}[h]
\includegraphics[width=0.95\linewidth]{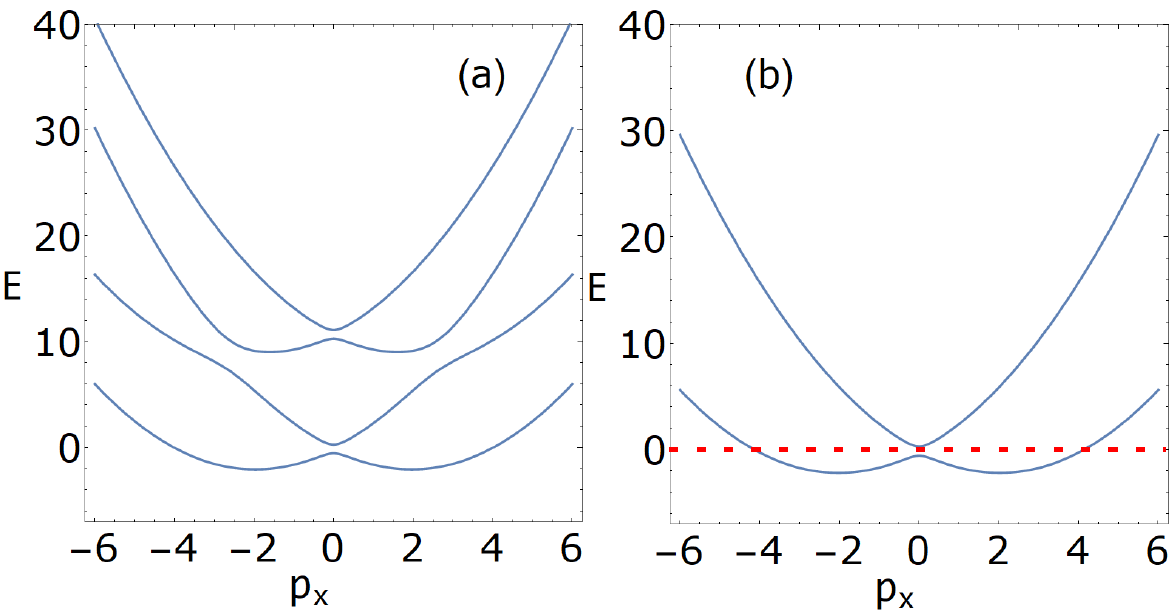}
\caption{(Color online) Plots of the energy spectra vs. momentum which demonstrate that the low-energy Hamiltonian we derived in Eq.~(\ref{eq:EffWire}) accurately reproduces the full four-band spectrum in the region around the lowest energy band. This allows us to analyze the topological transition (i.e. the gap opening and closing around $p_x=0$) using the more tractable $2 \times 2$ model. Panel (a) is the energy spectrum with finite interwire hopping and in-plane magnetic field including all four bands. All units and parameters are chosen to be same as Fig.~\ref{fig:Two}(c). Panel (b) is the energy spectrum of the projected Hamiltonian  $\mathcal{H}^{\mathrm{eff}}_{\Delta=0}$. \label{fig:Three}}
\end{figure}

Next we want to find an effective Hamiltonian that describes the two lowest-energy bands so we can analyze the topological properties of our model. We rearrange the basis of the $4 \times 4$ Hamiltonian in Eq.~(\ref{eq:HamNOdelta}) into $(\psi_{1\downarrow},\psi_{2\uparrow},\psi_{1\uparrow},\psi_{2\downarrow})^T$ such that the upper-left block corresponds to the Hilbert space that we want to project into, namely, the lowest two energy bands spanned by the states $(\psi_{1\downarrow},\psi_{2\uparrow})$:

\begin{widetext}
\begin{equation}
\mathcal{\widetilde{H}}_{\Delta=0}=
\left[ {\begin{array}{cccc}		
-B_z -\mu + \frac{(p_x-A_x)^2}{2m} & 0 & B_x & t \\
0 & -B_z -\mu + \frac{(p_x+A_x)^2}{2m} & t & B_x \\
B_x & t &B_z -\mu + \frac{(p_x-A_x)^2}{2m} & 0 \\
t & B_x &0 & B_z -\mu + \frac{(p_x+A_x)^2}{2m} \\
\end{array} } \right]~.\label{eq:Wire}
\end{equation}	
\end{widetext}
Now that we have written the Hamiltonian in a form
\begin{equation}
\mathcal{\widetilde{H}}_{\Delta=0}=
\left[
\begin{array}{c|c}
\mathcal{H}_{11} & \mathcal{H}_{12} \\
\hline
\mathcal{H}_{21} & \mathcal{H}_{22}
\end{array}
\right]~,\label{eq:Block}
\end{equation}
we can apply second-order perturbation theory to project the Hamiltonian onto the low-energy subspace of the two lowest energy bands. This procedure relies on the energy scales of the off diagonal Hamiltonians (i.e., $B_x$ and $t$) being much smaller than the energy scales of the diagonal Hamiltonians (i.e., $\pm B_z - \mu + \frac{1}{2 m} (p_x \pm A_x)^2$). The effective Hamiltonian in the low-energy subspace is given by
\begin{equation}
\mathcal{H}^{\mathrm{eff}}_{\Delta = 0} = \mathcal{H}_{11} \, + \, \mathcal{H}_{12}\left[{\mathcal{H}_{22}}\right]^{-1}\mathcal{H}_{21}~,\label{eq:ProjectionWire}
\end{equation}
which when written out explicitly, looks like
\begin{widetext}
\begin{equation}
\mathcal{H}^{\mathrm{eff}}_{\Delta=0}=
\left[ {\begin{array}{cc}
-B_z -\mu + \frac{(p_x-A_x)^2}{2m}  + \bigg(\frac{B_x^2 + t^2}{B_z -\mu + (p_x-A_x)^2/2m} \bigg)& \frac{2B_xt \, \big(B_z-\mu + (A_x^2 + p_x^2)/2m \big)}{\big(B_z -\mu + (p_x+A_x)^2/2m\big)\big(B_z -\mu + (p_x-A_x)^2/2m\big)}\\
\frac{2B_xt \, \big(B_z-\mu + (A_x^2 + p_x^2)/2m \big)}{\big(B_z -\mu + (p_x+A_x)^2/2m\big)\big(B_z -\mu + (p_x-A_x)^2/2m\big)}
& -B_z -\mu + \frac{(p_x+A_x)^2}{2m} + \bigg(\frac{B_x^2 + t^2}{B_z -\mu + (p_x+A_x)^2/2m}\bigg) 
\end{array} } \right]~.\label{eq:EffWire}
\end{equation}
The off-diagonal terms in Eq.~(\ref{eq:EffWire}) provide the desired effective coupling between spin-down states whose wave function is primarily in the first wire and spin-up states whose wave function is primarily in the second wire, which to leading order creates a gap at $p_x=0$ which is $\sim  \! t B_x/B_z$. The additional terms appearing on the diagonals $\sim \! (B_x^2 + t^2)/B_z$ can primarily be understood as a small correction to the global zero of energy. We compare the eigenvalue spectra of   $\mathcal{\widetilde{H}}_{\Delta=0}$ and $\mathcal{H}^{\mathrm{eff}}_{\Delta=0}$ in Fig. \ref{fig:Three}(a) and Fig. \ref{fig:Three}(b), respectively, which demonstrates that the projection accurately captures the spectra of the two lowest-lying bands in the energy window around the near degeneracy at $p_x=0$. 
\end{widetext}	

\section{Proximity induced $S$-wave superconductivity and emergent Majorana Zero Modes}
\label{Sect:Two}

In this section we will add intrawire and interwire superconducting pairing terms to our Hamiltonian. We take these terms to originate from proximity coupling to an $s$-wave superconductor on which the two wires are placed (see Fig.~\ref{fig:One}). And while we do not explicitly consider it in this paper, we note that it has been demonstrated that power-law superconducting order generated by interactions native to the quantum wire can, under certain circumstances, still yield Majorana quasiparticles at the wire's ends~\cite{fidkowski2011majorana,sau2011number}. We use the low-energy model derived in the previous section to analytically construct the MZM's wave function in the regime where interwire hopping and in-plane Zeeman energies are small compared to the out-of-plane Zeeman energy. We then relax the condition of small interwire hopping and small in-plane Zeeman coupling and we numerically calculate the topological phase diagram throughout different regions of parameter space using a Pfaffian formulation for the topological invariant. 

We begin by adding a momentum independent (i.e., $s$-wave) intrawire pairing potential, $\Delta_1$, and interwire pairing potential, $\Delta_2$, to obtain our full $8 \times 8$ Bogoliubov-de Gennes (BdG) Hamiltonian,
\begin{equation}\label{eq:hopping}
\begin{array}{l}
{\displaystyle  \! \! \! \! \! \! \! \! \! \! \mathcal{H}_{\rm BdG} = \left(\frac{p_x^2+A_x^2}{2m} - \mu \right)\tau_z }\vspace{0.2 cm}\\
{\displaystyle \quad \, - \left( \frac{A_xp_x}{m}\rho_z - B_z\sigma_z\rho_z - B_x\sigma_x - t\rho_x\right)\tau_z }\vspace{0.2 cm}\\
{\displaystyle \quad \, -\bigg(\Delta_1\sigma_y+\Delta_2\sigma_y\rho_x\bigg)\tau_y} ~,
\end{array}
\end{equation}
where our basis here is defined by
\begin{equation}
\psi=(\psi_{1\uparrow},\psi_{1\downarrow},\psi_{2\uparrow},\psi_{2\downarrow},\psi^\dagger_{1\uparrow},\psi^\dagger_{1\downarrow},\psi^\dagger_{2\uparrow},\psi^\dagger_{2\downarrow})^T ~,\label{eq:Hamiltonian}
\end{equation}
and where the Pauli matrices $\sigma, \rho$, and $\tau$ act on spin, wire, and particle-hole degrees of freedom, respectively. As in Eq.~(\ref{eq:HamNOdelta}), we have chosen the units here such that $e=c=\mu_B=\hbar=1$ and the Land\'e factors are equal in magnitude and can be absorbed into the components of the vector $\vec{B}$ for convenience.

One can define the anti-unitary charge conjugation and time-reversal operators, $\mathcal{C}$ and $\mathcal{T}$, as $\tau_xK$ and $i\sigma_yK$, respectively, where $K$ is the complex conjugation operator. The former satisfies $\{\mathcal{C},\mathcal{H}\}=0$ whereas the latter does not satisfy $[\mathcal{T},\mathcal{H}] =0$. This is expected, as the system contains magnetic field and hence the time-reversal invariance is explicitly broken. Since $\mathcal{C}^2 = 1$, our Hamiltonian $\mathcal{H}$ belongs to the \emph{D} class in the tenfold way\cite{ryu2010topological,schnyder2008classification}. In spatial dimension $d=1$, systems 
in this class carry a $\mathbb{Z}_2$ topological invariant.

\subsection{Majorana zero modes in the low-energy subspace defined by $B_x / B_z \ll  1$}
\label{Subsect:Two_A}
Just as we obtained an effective Hamiltonian describing the low-energy sector of our model in the absence of superconductivity (see Sec.~\ref{Sect:One}), we want to identify how $\Delta_1$ and $\Delta_2$ combine with the other parameters of our model to create a single effective pairing potential $\Delta^*$ which acts within the low-energy sector to create Cooper pairs from our helical fermions. We again start by reordering the $8 \times 8$ Hamiltonian $\mathcal{H}_{\rm BdG}$ of Eq.~(\ref{eq:Hamiltonian}) into a equivalent matrix $\mathcal{\widetilde{H}}_{\rm BdG}$ which acts on the basis
\begin{equation}
\psi_{\mathrm{eff}}=(\psi_{1\downarrow},\psi_{2\uparrow},\psi^\dagger_{1\downarrow},\psi^\dagger_{2\uparrow},\psi_{1\uparrow},\psi_{2\downarrow},\psi^\dagger_{1\uparrow},\psi^\dagger_{2\downarrow})^T~. \label{eq:psi_lowE}
\end{equation}
The Hamiltonian is now organized into a block form 
\begin{equation}
\mathcal{\widetilde{H}}_{\rm BdG}=
\left[
\begin{array}{c|c}
\mathcal{\widetilde{H}}_{{\rm BdG},11} & \mathcal{\widetilde{H}}_{{\rm BdG},12} \\
\hline
\mathcal{\widetilde{H}}_{{\rm BdG},21} & \mathcal{\widetilde{H}}_{{\rm BdG},22}
\end{array}
\right], \label{eq:BlockTwo}
\end{equation}
and we can again project onto the low-energy subspace by restricting ourselves to the limit in which the energy scales of the terms in off-diagonal blocks are small compared to the diagonal terms. We again apply Eq.~(\ref{eq:ProjectionWire}) to obtain the $4 \times 4$ BdG Hamiltonian for the low-energy subspace
\begin{equation}
\mathcal{H}^{\mathrm{eff}}_{\rm BdG} = \mathcal{\widetilde{H}}_{{\rm BdG},11} \, + \, \mathcal{\widetilde{H}}_{{\rm BdG},12}\left[{\mathcal{\widetilde{H}}_{{\rm BdG},22}}\right]^{-1}\mathcal{\widetilde{H}}_{{\rm BdG},21}~.
\end{equation}
We refer to Appendix \ref{Sect:AppendixA} for the explicit, and rather lengthy, form of the effective BdG Hamiltonian. 

We can now begin to analyze the topological phases of the effective model governed by $\mathcal{H}^{\mathrm{eff}}_{\rm BdG}$. The presence of a finite $\Delta_2$ guarantees that when the chemical potential is such that only the lowest energy band is occupied (i.e., $\mu \lesssim \mu_{\rm gap}$), then there is always a finite gap in this band at the Fermi momentum $\pm p_{F}$ [i.e., a pairing gap opens where the red dashed-line marking the Fermi energy intersects the lowest energy band in Fig.~\ref{fig:Three} (b)]. And the degeneracy between the lowest two energy bands which appears at $p_x = 0$ when either $B_x=0$ or $t=0$, is also gapped by a finite $\Delta_2$. Just as in the proposal using a single quantum wire with strong Rashba spin-orbit \cite{oreg2010helical}, we will demonstrate that a topological phase emerges when the gap at $p_x = 0$ is dominated by the spin-flip interwire hopping term ($\sim t B_x/ B_z$) instead of the superconducting pairing term. To demonstrate the emergence of MZMs in the topological phase, we expand $\mathcal{H}^{\mathrm{eff}}_{\rm BdG}$ around $p_x\sim0$ up to linear order to obtain $\mathcal{H}^{\mathrm{eff}}_{\rm BdG}\left(p_x \sim 0 \right)$. Both $\mathcal{H}^{\mathrm{eff}}_{\rm BdG}$ and $\mathcal{H}^{\mathrm{eff}}_{\rm BdG}\left(p_x \sim 0 \right)$ have highly complicated forms that are inconvenient to show explicitly; see Appendix \ref{Sect:AppendixA} for the details of the projection to and expansion of $\mathcal{H}^{\mathrm{eff}}_{\rm BdG}$. The resulting linearized effective Hamiltonian is re-parametrized to have the form
\begin{multline}
\! \! \! \! \!  \mathcal{H}^{\mathrm{eff}}_{\rm BdG}\left(p_x \sim 0 \right) = \vspace{0.5 cm} \\
\quad \quad  \left[ \! \! \!  {\begin{array}{cccc}
-u^*p_x-\mu^* & t^* &\Lambda p_x & -\Delta^* \\
t^* &	u^*p_x-\mu^* &  \Delta^* & \Lambda p_x \\
\Lambda p_x & \Delta^* & 	-u^*p_x+\mu^* &-t^* \\
-\Delta^*& \Lambda p_x & -t^* &	u^*p_x+\mu^* \\
\end{array} } \! \! \right]. \label{eq:LinearEff}
\end{multline}
In Appendix \ref{Sect:AppendixA}, we give the explicit definitions for the new parameters $\Delta^*, t^*$, and $u^*$. These new parameters represent the effective superconducting pairing potential ($\Delta^* \propto \Delta_2 + ... $), the effective interwire hopping term combined with a spin-flip ($t^* \propto B_x t/B_z + ... $) and the effective velocity which imitates the spin-orbit coupling strength in wires with Rashba coupling ($u^* \propto A_x/m + ...$), respectively. The term $\Lambda p_x$ comes from a higher order correction, namely a combination of the intra-wire coupling $\Delta_1$ and spin-flip term $ \propto B_x$, and for the parameter space investigated in this paper it satisfies $\Lambda \ll \Delta^*, t^*, u^*$. 

With the exception of the terms $\propto \! \Lambda p_x$, all the terms in our effective model Eq.~(\ref{eq:LinearEff}) have analogs to the terms which appear in models of single quantum wires with strong Rashba spin-orbit interactions (see e.g., \cite{oreg2010helical}). As such, we are motivated to find the analytic form of the Majorana wave function using the same procedure: We consider a spatial variation of $t^*$ that generates two regions where the gap at $p_x=0$ is dominated either by $t^*$ or by $\Delta^*$, and on the boundary between these two regions MZMs emerge. More explicitly, we use periodic boundary conditions and, set $t^* = \Delta^* + ax$ for a small connected interval around $x=0$ and $t^* = \Delta^* + a(l/2-x)$ for a second small connected interval around $x= l/2$. These two small intervals divide the circle of length $l$ into two regions with  $t^*>\Delta^*$ and $t^*<\Delta^*$, where $t^*$ is not varying. The necessity for the second interval stems from considering periodic boundary conditions, wherein the value of $t^*$ should match at $0$ and $l$. As a result we can easily demonstrate that a MZM is bound to the small interval where $t^*$ linearly varies. The zero energy MZM eigenvalue is found by setting $\mu^*=0$ and squaring the effective Hamiltonian 
\begin{multline}
\left[\mathcal{H}^{\mathrm{eff}}_{\rm BdG}\left(p_x \sim 0 \right) \right]^2 =
\left[ {\begin{array}{cccc}
\epsilon & iau^* &\beta & ia\Lambda \\
-iau^* &\epsilon &  ia\Lambda & -\beta  \\
\beta &-ia\Lambda & \epsilon&-iau^* \\
-ia\Lambda&-\beta & iau^*  &\epsilon \\
\end{array} } \right]\hfill
\label{eq:squared}
\end{multline}
where 
\begin{equation}
\epsilon = {\Delta^*}^2+{t^*}^2+{u^*}^2p_x^2+ \Lambda^2 p_x^2
\end{equation}
and $\beta = 2t^*\Delta^*-2p_x^2u$. The terms $iau^*$ and $ia\Lambda$ come from the fact that the coupling $t^*$ now has a spatial dependence, and due to the commutator $[x,p] = i$, these extra terms emerge. To identify the zero-energy modes we diagonalize Eq.~(\ref{eq:squared}) with a unitary transformation defined by the rotation matrix:
\begin{multline}\centering
\renewcommand{\arraystretch}{1.3}
\begin{aligned}
\qquad \qquad U& = \frac{1}{2}(i\sigma_z + \tau_y - i\sigma_y\tau_z -\sigma_y\tau_x ) \\
\qquad \qquad&  =
\begin{bmatrix*}[r]
\frac{i}{2} & -\frac{1}{2} & -\frac{i}{2} & -\frac{1}{2} \\
\frac{1}{2} & -\frac{i}{2} & -\frac{1}{2} & -\frac{i}{2} \\
\frac{i}{2} & -\frac{1}{2} & \frac{i}{2} & \frac{1}{2} \\
-\frac{1}{2} & \frac{i}{2} & -\frac{1}{2} & -\frac{i}{2} \\
\end{bmatrix*}~.
\end{aligned}\label{eq:rotation}
\end{multline}
After diagonalizing the Hamiltonian, two out of the four diagonal terms have the form of a simple harmonic oscillator, namely, $(u^*+\Lambda)^2p^2 + a^2x^2 \pm a(u^*+\Lambda)$, whereas the other two diagonal elements have more complicated forms and are not of interest to us. The corresponding eigenvalues are simply given by $E^2 = 2a(u^*+\Lambda)(n+1/2) \pm (u^*+\Lambda)a$. Depending on the sign of $a$, one or the other of the diagonals have a $n=0$ mode with zero energy. For any sign choice, we always have two intervals in which one of the diagonals accommodate the MZM. Although we keep the additional term $\Lambda p_x$ in Eq.~(\ref{eq:LinearEff}), it does not affect the qualitative nature of the zero modes. This effective term corresponds to a spin-triplet $p$-wave superconducting pairing term and only appears in the MZM wave function as a modification of the effective spin-orbit coupling term $u^*$. Quantitatively it increases (decreases) the spread of the MZM wave function for $\Lambda>0 \, (\Lambda<0)$ as can be seen below.

The explicit form of the MZM can be found by applying the inverse transformation $U^{-1}$ on the eigenvectors of diagonalized matrix. Let $a>0$ and consider the MZM at $x=0$, which corresponds to the eigenvector $(0\, 1\, 0\, 0)^T$ in the diagonalized basis. Applying the inverse transformation gives the following Majorana operator,
\begin{equation}
\gamma = \gamma^\dagger = 1/2(-\psi_{1\downarrow} -i\psi_{2\uparrow} -\psi^\dagger_{1\downarrow} +i\psi^\dagger_{2\uparrow})\label{eq:Majorana}
\end{equation}
which satisfies the defining property of a Majorana mode. The spatial extent of the MZM is given by the well-known $n=0$ solution of the simple harmonic oscillator $\Psi(x) \propto \exp{\left[-ax^2/(2 u^*-2 \Lambda) \right]}$.

\subsection{Majorana Zero Modes for $B_x / B_z \gtrsim 1$ }
\label{Subsect:Two_B}
In the last section we have identified the conditions for the emergence of MZMs when our low-energy model is valid (i.e., when the energy scales associated with spin-conserving interwire hopping and in-plane Zeeman coupling are much smaller than the out-of-plane Zeeman coupling energy scale). We can now ask whether or not a topological phase yielding emergent MZMs can exist outside of the parameter space where the low-energy model is valid. By numerically evaluating the $\mathbb{Z}_2$ topological invariant by using a Pfaffian representation~\cite{kitaev2001unpaired,lutchyn2011search}, we will answer this question in the affirmative. 

We follow the procedure of Ref.~\cite{lutchyn2011search} and look at the Pfaffian of the skew-symmetric matrix  $\mathcal{H}_{\rm BdG}\tau_x$, which has the same set of eigenvalues. The Pfaffian of a skew-symmetric matrix gives the square root of the determinant, which can be thought of as multiplication of half of the eigenvalues~\cite{ryu2007z}. The change in sign of the Pfaffian indicates a gap opening and closing transition implying a band inversion has occurred. This allows one to build a topological invariant by comparing the signs of the Pfaffian at $p_x=0$ and $p_x \rightarrow \infty$~\cite{kitaev2001unpaired}. The system is in the topological superconducting phase and supports end-state MZMs when the Pfaffian has opposite signs at these given points, 
\begin{equation}
{\rm Pf}\big[\mathcal{H_{\rm BdG}} (p_x = 0)\tau_x\big] \times {\rm Pf}\big[\mathcal{H_{\rm BdG}} (p_x \rightarrow \infty )\tau_x\big] = -1 ~, \label{eq:TopInv}
\end{equation}
and is in the trivial superconducting phase when the left-hand side of Eq.~(\ref{eq:TopInv}) equals $+1$. We calculated the Pfaffian for different $p_x$ values by applying a freely available numerical package~\cite{wimmer3440efficient}. We found that for a finite pairing potential $\Delta_2$ and for $\mu \sim \mu_{\rm gap}$ the Pfaffian is positive definite as $p_x \rightarrow \infty$. Hence, it is sufficient to look at only the sign of ${\rm Pf}[\mathcal{H}(p_x=0)\tau_x]$. This can be calculated analytically and is given by

\begin{multline}
\begin{aligned}
&{\rm Pf}\bigg[\mathcal{H}_{\rm BdG}(p_x=0)\tau_x\bigg]= \frac{A_x^8}{16 m^4} -\frac{A_x^6\mu}{2m^3} \\
&\qquad \quad \,\,\,\, -\frac{A_x^4}{2m^2}(B_x^2+B_z^2+t^2-\Delta_1^2-\Delta_2^2-3\mu^2)\\
&\qquad \quad+\frac{2A_x^2}{m}\Big((B_x^2+B_z^2)\mu-2t\Delta_1\Delta_2+ t^2\mu\Big)\\
&\, \, \, \,\, \, \, \,\, \, \, \,-\frac{2A_x^2}{m}\Big(\mu(\Delta_1^2+\Delta_2^2+\mu^2)\Big)\\
&\, \, \, \,\, \, \, \,+\Big(2B_z^2(t^2-\Delta_1^2+\Delta_2^2-\mu^2)\\
&\, \, \, \, +B_x^4+B_z^4+2B_x^2(B_z^2-t^2-\Delta_1^2-\Delta_2^2-\mu^2) \\
&+[(\Delta_1^2+\Delta_2^2)^2+ (t-\mu)^2][(\Delta_1^2-\Delta_2^2)^2 + (t+\mu)^2]\Big).\label{eq:Pfaffian}
\end{aligned}
\end{multline} 

\begin{widetext}

\begin{figure}[h!]
\includegraphics[width=0.95\linewidth]{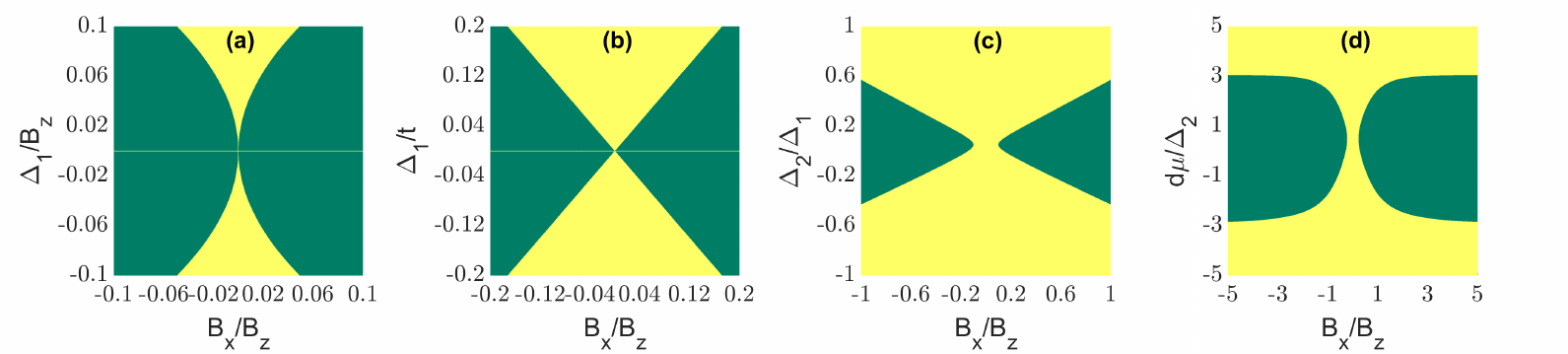}
\caption{(Color online)  Phase diagrams of the general Hamiltonian $\mathcal{H}_{BdG}$ given in Eq.~(\ref{eq:Hamiltonian}). Green (yellow) regions indicate the parameter values where system is (not) topological, i.e. $Pf[\mathcal{H}(p_x=0)] <(\geq) 0$. Explicit definitions of the parameters are given in the main text. Panel (a) compares the intra-wire SC coupling $\Delta_1$ and in-plane magnetic field $B_x$, in which both are scaled with respect to transverse magnetic field $B_z$. The other parameters are set to $\Delta_2/\Delta_1= 0.1$, $t/B_z = 0.1$, $A_x = \sqrt{0.16 B_z}$, and $\delta\mu = 0$. Panel (b) is plotted with the parameters $\Delta_2/\Delta_1 = 0.1$, $t/B_z = 1$, $A_x = \sqrt{0.16 B_z}$, $\delta\mu = 0$, and now $\Delta_1$ is scaled with respect to the interwire hopping amplitude, $t$. Panel (c) compares the strength of the in-plane magnetic field with ratio of the interwire, $\Delta_2$, and intra-wire, $\Delta_1$, SC couplings where $\Delta_1/B_z = 0.1$, $t/B_z = 0.05$, $A_x = \sqrt{0.16 B_z}$, and $\delta\mu = 0$. Finally, panel (d) compares the in plane magnetic field and variation in chemical potential, $\delta \mu$, with respect to interwire SC coupling with $\Delta_2/B_z = 0.01$, $\Delta_1/B_z = 0.1$, $t/B_z = 0.03$, and $A_x = \sqrt{0.16 B_z}$. \label{fig:Four}}
\end{figure}

\end{widetext}

In Fig.~\ref{fig:Four}, we present the topological phase diagram of our system obtained using the Pfaffian formulation of the topological invariant. As the parameter space is vast, we limit our illustrations to cross-sections of this space that yield some important characteristics. In all panels of Fig.~\ref{fig:Four}, the coupling $A_x$ is scaled by $A_x \rightarrow A_x\sqrt{m}$ and the chemical potential is set to $\mu = \mu_{\mathrm{gap}} + \delta\mu$ where the first term is defined in Eq.~(\ref{eq:mu}), and the second term describes variations. The yellow (green) regions are where the system is trivial (topological) and on the boundary between these regions a gap closing and reopening takes place at $p_x=0$ as the system undergoes a topological phase transition.

The four panels of Fig.~\ref{fig:Four} reveal a number of interesting results for us to discuss. Figure~\ref{fig:Four}(a)  confirms the presence of the topological phase which we previously discovered in Sec.~\ref{Sect:Two} wherein we derived an effective low-energy $4 \times 4$ BdG Hamiltonian which was valid in the limit of small $B_x$ and $t$. It is reassuring that the MZMs are also found when the topological invariant is calculated by calculating the Pfaffian of the full $8 \times 8$ BdG Hamiltonian. Note that the thin horizontal yellow line in Figs.~\ref{fig:Four}(a) and ~\ref{fig:Four}(b) is nothing other than a symptom of our choice to fix $\Delta_2 = 0.1 \Delta_1$ and $\Delta_2$ must be finite for the gap in the lowest energy band (away from $p_x = 0$) to be finite. In Fig.~\ref{fig:Four}(b), we consider a large interwire spin-conserving hopping term, $t = B_z$, which is outside of the regime of validity for the low-energy model we derived using second-order perturbation theory, but yet a topological phase transition is still present. This is likely due to the fact that, although our perturbation theory for the low energy subspace breaks down and we cannot use $\mathcal{H}^{\mathrm{eff}}_{\rm BdG}$, the physics of the topological phase transition is still governed by a competition between two gapping sources of the near-degeneracy at $p_x=0$, and indeed the topologically non-trivial region in Fig.~\ref{fig:Four}(b) still satisfies $2 \Delta_2 \lesssim t B_x/B_z$. We find in Fig.~\ref{fig:Four}(c) that when $B_x<2B_z$ the interwire pairing potential should be less than the intrawire pairing potential for a non-trivial phase to emerge; $\Delta_2 < \Delta_1$ is indeed the case in almost all real world systems. Finally in Fig.~\ref{fig:Four}(d) we demonstrate that there is an almost constant window for varying the chemical potential, i.e., $\delta \mu \neq 0$, which can be interpreted as some measure of the gap in units of $\Delta_2$. This range appears to be linearly proportional to the relative strength of $t$ but independent of $B_x$. Also, we observe that the topological phase persists for even large values of in-plane Zeeman coupling, $B_x/B_z > 1$.
\section{Emergent Fractional Excitations}
\label{Sect:Three}
Ever since Kane, {\it et al.}~\cite{kane2002fractional} demonstrated that electron-electron interactions in a two-dimensional-array of tunnel-coupled quantum wires can yield fractionally-charged excitations analogous to the anyonic excitations appearing in fractional quantum Hall phases of two-dimensional electron gases, numerous studies have applied the same approach for generating topological phases and fractionalized quasiparticles~\cite{hu2018fibonacci,strinati2017laughlin,bulmash2017strongly,iadecola2016wire,sagi2015array,meng2015coupled,klinovaja2014time}. In this section we incorporate electron-electron interactions into our model using bosonization techniques. We then discuss how, and for which specific regions of our model's parameter space, these interactions generate fractional zero modes that obey $\mathbb{Z}_3$ parafermionic algebra in place of the MZMs which appear in the non-interacting system.

We begin by considering our model, Eq.~(\ref{eq:HamNOdelta}), when the density of electrons has been tuned by external electrostatic gates so {\it only} the lowest energy band in Fig.~\ref{fig:Two}(c) is occupied. More specifically, we set the chemical potential such that the Fermi energy lies below the gap separating the lowest and second-lowest energy bands. We restrict ourselves to considering electron-electron interactions whose strength, $U$, is much less than the energy difference, $\Delta E_{\rm band}$, of the lowest energy band between its two degenerate band minima and the gap to the second band. When the temperature is also significantly less than this bandwidth ($k_{\rm B} T \ll \Delta E_{\rm band}$), the effective Hilbert space of our model is then reduced to the electronic states within the lowest energy band which are within a small energy window near the four Fermi points $\pm k_{\rm F} \pm k_A$, where $k_A$ is the shift in the fermi wavevector due to orbital coupling. We can approximate the dispersion of the lowest energy band (when the chemical potential is close to the doubly degenerate band-edge) as two parabola, $(p_x \pm A_x)^2/2m$, and introduce a Fermi wavevector to define the density of electrons in each parabola, $k_F = \rho_{i=1,2}/\pi$. Since the fermions in the lowest energy band are helical [see Figure~\ref{fig:Two} (d)], the states in the two parabola have opposite spin. The resulting model's non-interacting dispersion is then identical to Ref.~\cite{oreg2014fractional}. Using the standard bosonization scheme we linearize the spectrum at the four points, $\pm k_{\rm F} \pm k_A$, which gives us two right-movers and two left-movers, one for each spin channel (see Fig. \ref{fig:Six} in Appendix \ref{Sect:AppendixB}).

In terms of the traditional spin-resolved charge-density and current-density fields~\cite{giamarchi2004quantum}, $\nabla \phi_{\uparrow(\downarrow)}$ and $\nabla \theta_{\uparrow(\downarrow)}$, respectively, the low-energy Hamiltonian can be written as
\begin{equation}
\mathcal{H}_{B} \!=\! \! \sum_{i = C,S}  \int \! \frac{dx}{2\pi} \, \, \Big[  u_i \kappa_i \left( \nabla \theta_i \right)^2 + \frac{u_i} {\kappa_i} \left( \nabla \phi_i \right)^2 \! \Big],
\label{eq:BosonHam}
\end{equation}
where $\theta_{C (S)} = (\theta_{\uparrow} \pm \theta_{\downarrow} )/\sqrt{2}$ and $\phi_{C (S)} = (\phi_{\uparrow} \pm \phi_{\downarrow} )/\sqrt{2}$. The definitions for the velocities $u_i$, and Luttinger parameters, $\kappa_i$, can be found in Ref.~\cite{giamarchi2004quantum}. Note that in Eq.~(\ref{eq:BosonHam}) we have not yet accounted for any proximity-induced superconducting pairing terms. And although Eq.~(\ref{eq:BosonHam}) accounts for most types of electron-electron scattering processes, it neglects any terms which cannot be written as quadratic powers of the $\theta$ and $\phi$ fields.

Electron interaction terms which cannot be written as quadratic powers of the $\theta$ and $\phi$ fields will generally appear in the Hamiltonian as cosines acting on linear combinations of these fields. Sources of these terms include spin-flip scattering, umklapp backscattering, superconducting pairing processes, and $(N\! > \!2)$-particle scattering processes. Whether or not these terms lead to spontaneous formation of gaps at the Fermi energy can be determined from a renormalization group (RG) analysis~\cite{giamarchi2004quantum}. Following Ref. ~\cite{oreg2014fractional} we include the following three-body backscattering term
\begin{equation}
\mathcal{O}^{\mathrm{BS}}_{3b} = g^{\mathrm{BS}}_{3b}\,   \left[ (\psi^\dagger_{L\downarrow}\psi_{R\downarrow}) (\psi^\dagger_{L\downarrow}\psi_{R\uparrow}) (\psi^\dagger_{L\uparrow}\psi_{R\uparrow}) + \mathrm{H.c.}\right] \label{eq:Inthopp}
\end{equation}
which requires special parameters of the model to be momentum conserving. In our case we can enforce a momentum conservation condition ($A_x = 3k_F$) by adjusting either the total electron density (as in the case of Ref.~\cite{oreg2014fractional}) or by adjusting the distance separating the two wires, $y_0$, or the magnitude of the external field, $B_z$.

One can also consider the effect of spin-flip interwire hopping $t^*$ in Eq.~\eqref{eq:LinearEff}, which gives the following term
\begin{equation}
\mathcal{O}_{t^*} =2t^* \! \!\! \sum_{\substack{\tau,\tau' = R,L\\\sigma = \uparrow,\downarrow}} \left[\psi^\dagger_{\tau,\sigma}\psi_{\tau',-\sigma}e^{-i\sigma(6 + \tau -\tau')k_Fx} +\mathrm{H.c.}\right],
\end{equation}
where $\tau,\tau'$ is $+1$$(-1)$ for $R$$(L)$, and $\sigma$ is $+1 (-1)$ for $\uparrow \!(\downarrow)$ in the exponent. This term is suitable within the assumptions made in Section~\ref{Sect:Two} where we derived a low-energy non-interacting Hamiltonian for the subspace spanned by the two lowest-energy bands. When written in the bosonization language we obtain eight integrals over terms $\cos(\varphi_{\tau',-\sigma}-\varphi_{\tau,\sigma} -\sigma(6 + \tau -\tau')k_Fx)$. This term averages to zero when the integral over $x$ is carried out because the $\varphi$ fields vary much slower then $k_Fx$. 

Next we will incorporate the effects of superconducting pairing terms which are enabled by placing our double-wire setup in proximity to an $s$-wave superconductor. From Eq.~(\ref{eq:LinearEff}) we identify a term proportional to $\Delta^*$ which looks like a spin-singlet $s$-wave pairing potential, and a term proportional to $\Lambda p_x$ which looks like a spin-triplet $p$-wave pairing potential. For the latter we obtain the following term:
\begin{equation}
\mathcal{O}^{\mathrm{SC}}_{trip.} = \Lambda \! \! \sum_{\substack{\tau = R,L \\ \sigma = \uparrow,\downarrow}} \,  \left[ \psi^\dagger_{\tau,\sigma} p_x \psi^\dagger_{\tau,\sigma}e^{-i2(3\sigma-\tau)k_Fx}  +\mathrm{H.c.} \right], \label{eq:SCtriplet}
\end{equation}
where we again note that $\Lambda$ now denotes the renormalized parameter which is appropriate to the energy scale at which linearization of the non-interacting spectrum has been taken. When written in the bosonization language, now we obtain four integrals over terms $\cos(2\varphi_{\tau,\sigma} + 2(3\sigma-\tau)k_Fx)$ which again averages to zero when the integral over $x$ is carried out. Turning our attention to the spin-singlet $s$-wave pairing term we find
\begin{equation}
\mathcal{O}^{\mathrm{SC}}_{\mathrm{sing.}} =2\Delta^*\! \!\! \sum_{\substack{\tau,\tau' = R,L}} \left[\psi^\dagger_{\tau,\uparrow}\psi^\dagger_{\tau',\downarrow}e^{-i2(1+\tau\tau')k_Fx} +\mathrm{H.c.}\right]. \label{eq:SCsinglet}
\end{equation}
Out of the four terms in this sum, only two of them have vanishing exponential, when $\tau= -\tau'$, and must therefore be analyzed within the renormalization group analysis we pursue below. Similar to $\Lambda$ which appears in Eq.~(\ref{eq:SCtriplet}), $\Delta^*$ here represents the renormalized value of the parameter which appears in Eq.~(\ref{eq:LinearEff}). Finally, in addition to the three-body term given in Eq.~(\ref{eq:Inthopp}), we also consider a momentum conserving three-body scattering process which includes superconducting pairing
\begin{equation}
\mathcal{O}^{\mathrm{SC}}_{3b} = g^{\mathrm{SC}}_{3b}\!  \left[ \psi^\dagger_{L\uparrow} (\psi^\dagger_{L\uparrow}\psi_{R\uparrow}) (\psi_{L\downarrow}\psi^\dagger_{R\downarrow}) \psi^\dagger_{R\downarrow} + \mathrm{H.c.} \right]
\label{eq:SC3body}
\end{equation}
We can write Equations~(\ref{eq:Inthopp}), (\ref{eq:SCsinglet}) and (\ref{eq:SC3body}) in the bosonization framework using $\psi_{(R/L)(\uparrow/\downarrow)} \sim e^{i\varphi_{(R/L)(\uparrow/\downarrow)}}$ and the transformation
\begin{equation}
\renewcommand{\arraystretch}{1.6}
\left( \! \! {\begin{array}{cccc}\phi_C \\ \phi_S \\ \theta_C \\  \theta_S	\end{array} } \! \! \right) =\begin{bmatrix*}[r]
\frac{1}{2} & -\frac{1}{2} &\frac{1}{2} & -\frac{1}{2}\\
\frac{1}{2} &-\frac{1}{2} &-\frac{1}{2} & \frac{1}{2}  \\
\frac{1}{2}&\frac{1}{2} & \frac{1}{2}&\frac{1}{2}\\
\frac{1}{2}&\frac{1}{2}& -\frac{1}{2}  &-\frac{1}{2} \\
\end{bmatrix*}  \left( \! \! {\begin{array}{cccc}\varphi_{L\uparrow} \\\varphi_{R\uparrow} \\ \varphi_{L\downarrow} \\\varphi_{R\downarrow} \end{array} } \! \! \right) \label{eq:TransformMat}~.
\end{equation}
We obtain in the new basis
\begin{equation}
\renewcommand{\arraystretch}{1.6}
\left\{ \! \! {\begin{array}{ccc} \mathcal{O}^{\mathrm{BS}}_{3b} \\ \mathcal{O}^{\mathrm{SC}}_{\mathrm{sing.}} \\ 	\mathcal{O}^{\mathrm{SC}}_{3b} \end{array} } \! \! \right\} =  \left\{ \! \! {\begin{array}{ccc}g^{\mathrm{BS}}_{3b}\cos\Big(3\phi_C -\theta_S \Big)  \\ \Delta^*\big[\! \cos(\theta_C - \phi_S) + \cos(\theta_C + \phi_S)\big] \\ g^{\mathrm{SC}}_{3b}\cos\Big(3\phi_S +\theta_C  \Big) \end{array} } \! \! \right\} \label{eq:IntOpCan}
\end{equation}
and we can now write down explicitly the full Hamiltonian which governs the phase diagram of our double-wire system when in proximity to an $s$-wave superconductor,
\begin{equation}
\begin{aligned}
\mathcal{H}_{B + SC} \Big\vert_{A_x = 3k_F}\! \! \! \!&=\! \! \sum_{i = \sigma,\rho}  \int \! \frac{dx}{2\pi} \, \, \Big[  u_i \kappa_i \left( \nabla \theta_i \right)^2 + \frac{u_i} {\kappa_i} \left( \nabla \phi_i \right)^2 \! \Big]  \vspace{0.5cm}\\
&\, \, \, \, + \int dx \,\, \left[ \mathcal{O}^{\mathrm{BS}}_{3b} + \mathcal{O}^{\mathrm{SC}}_{\mathrm{sing.}} + \mathcal{O}^{\mathrm{SC}}_{3b} \right]~. \label{eq:fullHamBoson}
\end{aligned}
\end{equation}
We want to demonstrate that the ground-state wave function of this Hamiltonian can possess fractional excitations in certain circumstances (i.e., in the strongly correlated regime). To obtain fractional zero modes, we require the spectrum to be fully gapped. The terms in the first line of Eq.~(\ref{eq:fullHamBoson}) describe a gapless and transitionally invariant system with linearly dispersive charge and spin excitations.  However, each cosine term in the second line of Eq.~(\ref{eq:fullHamBoson}) will try to pin the combination of fields which appear in its arguments to the value $ \pi \,{\rm mod} \,2\pi$, and when the coupling constants for these terms, $(g^{\mathrm{BS}}_{3b}, \Delta^*, g^{\mathrm{SC}}_{3b} )$, are sufficiently large, the charge and/or spin excitation spectrum can become gapped.  Whether or not this happens depends on the material-specific values of the coupling constants $(g^{\mathrm{BS}}_{3b}, \Delta^*, g^{\mathrm{SC}}_{3b} )$ and whether they grow (i.e., are relevant operators) as we follow the renormalization group flow to long wavelengths. We follow the discussion in Refs.~\cite{oreg2014fractional,stoudenmire2011interaction,klinovaja2014parafermions}, and aim simply to argue that the fully gapped state which can support fractional excitations is not precluded.

Since $\mathcal{O}^{\mathrm{BS}}_{3b}$ and $\mathcal{O}^{\mathrm{SC}}_{\mathrm{sing.}}$ do not commute with each other, only one or the other can be good quantum numbers of the system, but neither one can fully gap both the charge and spin excitations of the system on their own. The terms $\mathcal{O}^{\mathrm{BS}}_{3b}$ and $\mathcal{O}^{\mathrm{SC}}_{3b}$, however, do commute and as we shall show below, they fully gap the bulk excitation spectrum while simultaneously creating emergent fractional zero modes at the ends of the wires. Since the bare values of the  model [i.e., the value of $(g^{\mathrm{BS}}_{3b}, \Delta^*, g^{\mathrm{SC}}_{3b} )$] get renormalized as we consider lower-energy and longer-wavelength scales, we must find which interaction grows during the process. To do this we calculate the scaling dimensions for each interaction in terms of Luttinger interaction parameters ($\kappa_S,\kappa_C$) for the spin and charge channels. As is common for the sine-Gordon equation, the scaling dimensions of the operators, $(\mathcal{O}^{\mathrm{BS}}_{3b}, \mathcal{O}^{\mathrm{SC}}_{\mathrm{sing.}} ,\mathcal{O}^{\mathrm{SC}}_{3b} )$, directly lead to the one-loop renormalization group equations for the coupling constants, $(g^{\mathrm{BS}}_{3b}, \Delta^*, g^{\mathrm{SC}}_{3b} )$, respectively~\cite{giamarchi2004quantum}. We find the scaling dimensions to be
\begin{equation}
\renewcommand{\arraystretch}{1.6}
\left\{ \! \! {\begin{array}{ccc} \gamma^{\mathrm{BS}}_{3b}   \\ \gamma^{\mathrm{SC}}_{\mathrm{sing.}} \\ 	\gamma^{\mathrm{SC}}_{3b} \end{array} } \! \! \right\} =  \left\{ \! \! {\begin{array}{ccc}  (9\kappa_C +\kappa^{-1}_S)/2 \\ (\kappa^{-1}_C + \kappa_S)/2 \\ (\kappa^{-1}_C + 9\kappa_S)/2  \end{array} } \! \! \right\} \label{eq:Scaling}
\end{equation}
These demonstrate that it is not possible for a system to have parameters $\kappa_C$ and $\kappa_S$ which simultaneously make $\gamma^{\mathrm{BS}}_{3b} < 1$ and $\gamma^{\mathrm{SC}}_{3b} <1$, which is required for these interactions to both flow to large coupling values regardless of how arbitrarily small the initial {\it bare} values of the coupling constants, $(g^{\mathrm{BS}}_{3b}, g^{\mathrm{SC}}_{3b} )$, are assumed to be. If one of $(\mathcal{O}^{\mathrm{BS}}_{3b}, \mathcal{O}^{\mathrm{SC}}_{3b} )$ is relevant, then the other is irrelevant in the renormalization group sense. Luckily, we are dealing with finite length wires, and therefore even if a operator is irrelevant in the RG sense, if it starts with a large enough bare value then it may still remain fairly large at the length scales similar to the wire length, and can thus still gap the bulk system.

\begin{figure}[H]
\includegraphics[width=0.95\linewidth]{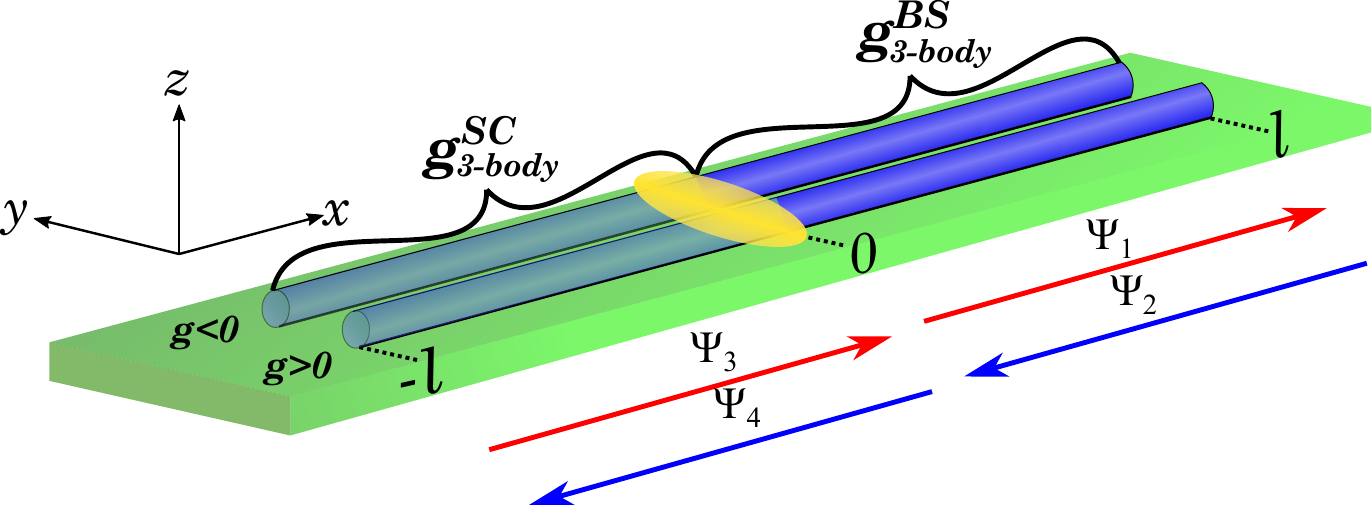}
\caption{(Color online)  Schematic representation of the unfolding scheme. Using the open boundary conditions allows \emph{unfolding} one of the right or left-mover fields into the virtually extended part of the wire (hallow). The process effectively doubles the size of the system. As a result interactions are confined within opposite regions and opens gaps of different nature. At the interfaces, which are the ends of the original wire, localized fractional zero modes emerge(yellow). \label{fig:Five}}
\end{figure}

Now we can consider what zero modes exist at the boundary in the case of a finite-length system in the strong-coupling limit where $\mathcal{O}^{\mathrm{BS}}_{3b}$ and $\mathcal{O}^{\mathrm{SC}}_{3b}$ pin the combination of fields in the argument of their cosines to $ \pi \,{\rm mod} \,2\pi$. We make the following transformation of fields 
\begin{equation}
\renewcommand{\arraystretch}{1.6}
\left\{ \! \! {\begin{array}{cccc} 	\Psi_1  \\  \Psi_2 \\ \Psi_3  \\  \Psi_4 \end{array} } \! \! \right\} =  \left\{ \! \! {\begin{array}{cccc}  \frac{1}{2}(\theta_C-3\phi_C +\theta_S-3\phi_S)  \\  \frac{1}{2}(\theta_C+3\phi_C -\theta_S-3\phi_S) \\ \frac{1}{2}(\theta_C+3\phi_C +\theta_S+3\phi_S) \\ \frac{1}{2}(\theta_C-3\phi_C -\theta_S+3\phi_S)  \end{array} } \! \! \right\} \label{eq:SimpleInt}
\end{equation}
such that the interactions simplify to
\begin{equation}
\renewcommand{\arraystretch}{1.6}
\left\{ \! \! {\begin{array}{cc} \mathcal{O}^{\mathrm{BS}}_{3b}   \\ 	\mathcal{O}^{\mathrm{SC}}_{3b} \end{array} } \! \! \right\} =  \left\{ \! \! {\begin{array}{cc}  g^{\mathrm{BS}}_{3b}cos\Big(\Psi_1 - \Psi_2 \Big)  \\  g^{\mathrm{SC}}_{3b}cos\Big(\Psi_3 + \Psi_4  \Big)  \end{array} } \! \! \right\}~. \label{eq:SimpleInt2}
\end{equation}
To uncover the presence of fractional zero modes at the ends of the wire governed by Eq.~(\ref{eq:SimpleInt2}) we apply the open boundary conditions via the well-known unfolding scheme~\cite{giamarchi2004quantum,fabrizio1995interacting}. In the spinless single-band quantum wire case, open boundary conditions demand that $\varphi_L(x) = \varphi_R(-x)$ for $0 \leq x \leq l$, which in turn allows us to double the size of the wire ($\left[0,l\right] \rightarrow \left[-l,l\right]$) while \emph{unfolding} either the right or left movers into the new extended part. Therefore, the finite length wire with two fields can be described after unfolding by a wire of doubled length in which each half contains only right-movers or only left-movers. We use this fact to enforce the following boundary conditions in our own setup:
\begin{equation}
\Psi_1(x) = \Psi_3(-x) \quad \Psi_2(x) = \Psi_4(-x) \quad 0 \leq x \leq l.
\end{equation}
Using this identification, we can define two fields in total for the extended wire
\begin{equation}
\Phi_{R(L)}(x) = \begin{cases} 
\Psi_{1(2)}(x) & 0\leq x\leq l\\
\Psi_{3(4)}(-x) & -l\leq x\leq0 
\end{cases}\label{eq:twofields}
\end{equation}
Comparing with Eq.(\ref{eq:SimpleInt2}) reveals that the two interactions which are required to fully gap the bulk of the double-wire setup each live within only one half of the unfolded setup of double the length (see Fig~\ref{fig:Five}). Because each half of the wire is gapped by a different interaction, zero energy modes can possibly emerge at the boundary. This is the interacting version of the statement that a system governed by the massive Dirac Hamiltonian supports zero energy modes where the sign of the mass changes~\cite{bernevig2013topological}.

Going back to the canonical fields, pinning gives the conditions
\begin{equation}
3\phi_C - \theta_S = \pi + 2\pi \hat{m} \quad {\rm and} \quad 3\phi_S + \theta_C = \pi + 2\pi \hat{n}~,
\end{equation}
where $\hat{m}$ and $\hat{n}$ are integer valued operators. Following Ref.~\cite{clarke2013exotic}, we calculate the commutation relation for $\hat{m}$ and $\hat{n}$. Using $[\phi(x),\theta(x')] = i\pi\Theta(x-x')$ and the fact that the domain of $\hat{m}$ is always to the right of the domain of $\hat{n}$, we find
\begin{multline}
\begin{aligned}
\pi^2[\hat{m},\hat{n}] &= 9[\phi_C,\phi_S] + 3([\phi_C,\theta_C] - [\theta_S,\phi_S]) - [\theta_S,\theta_S] \\
&= i3\pi \implies [\hat{m},\hat{n}] = \frac{i3}{4\pi}
\end{aligned}
\end{multline}
We then define the operator $\alpha = {\rm exp}\left[i2\pi(\hat{m}+\hat{n})/3\right]$ which commutes with the Hamiltonian at the boundary of two region, i.e., an infinitesimally small region where the Hamiltonian contains only the quadratic in $\theta_{C(S)}$ and $\phi_{C(S)}$ terms. Moreover we have $\alpha^3 = 1$ which satisfies the $\mathbb{Z}_3$ parafermionic algebra \footnote{However, it should be noted that parafermionic algebra also includes fractional statistics between different parafermion operators, i.e., $\alpha^{\dagger}_i\,\alpha_j = e^{-(2\pi i/3)\mathrm{sgn}(i-j)}\alpha_j\,\alpha^{\dagger}_i$ for two different parafermion modes $\alpha_{i,j}$\cite{clarke2013exotic}. The terminology we use is therefore adopted from parafermionic excitations in two dimensional systems.  As we have a single fractional mode in one dimension, fractional statistics condition is not applicable. See Refs. \cite{klinovaja2014parafermions,klinovaja2014time} for similar termionology in one dimension.}. Hence, in the neighborhoods of the interface between the regions gapped by $\mathcal{O}^{\mathrm{SC}}_{3b}$ and gapped by $\mathcal{O}^{\mathrm{BS}}_{3b}$, we establish the existence of emergent fractional zero modes in our particular setup.

We note that these fractional zero modes are 
not stable against local perturbations, i.e.,
the ground state degeneracies higher than two are
not topologically protected and, thus, can be lifted 
by disorder, as discussed in Refs.\ 
\cite{fidkowski2011topological,turner2011topological,
bultinck2017fermionic,oreg2014fractional,calzona2018z4}. 
Therefore, a highly clean 
and fine-tuned setup is needed to minimize disorder. 
Alternatively, 
stacking many copies of the proposed one-dimensional setup
into an effectively two-dimensional setup may 
suppress the lifting of the degeneracies as 
recently shown in the Ref.\ \cite{strinati2019pretopological}.

We conclude this section with a brief discussion of how the parameters of our specific setup can be best chosen to yield fractional modes. The most important condition is to meet the momentum conservation condition $k_A = 3k_F$. When $k_A - 3k_F = \delta$ the interaction strength $g^{\mathrm{BS}}_{3b}$ is reduced by a factor $\sim {\rm exp}(-\delta \ell)$, where $\ell$ is the length of the wires. Strain and external fields can impact the effective value of the Land\'e $g$ factors~\cite{winkler2017orbital}, which together with the magnetic field strength determines $k_A$, as discussed in Section~\ref{Sect:Two}. Since $\mathcal{O}^{\mathrm{SC}}_{3b}$ and $\mathcal{O}^{\mathrm{BS}}_{3b}$ cannot simultaneously be made relevant in an RG sense, it is important that at least one of them has a large bare coupling constant, $g^{\mathrm{SC}}_{3b}$ and $g^{\mathrm{BS}}_{3b}$, respectively. These coupling constants are $g^{\mathrm{SC}}_{3b}\sim \Delta^* V_{2k_F}$ and $g^{\mathrm{BS}}_{3b}\sim t^{*} V^2_{2k_F}$, respectively, where $V_{2k_F}$ is the Fourier transformation of the electron-electron interaction in one-dimension evaluated at $q = 2k_F$ \cite{giamarchi2004quantum}. See Appendix \ref{Sect:AppendixA} for the parameters $\Delta^{*}$ and $t^{*}$ expressed in terms of the original parameters of the model. In general, unless the number of electrons per unit cell approaches an integer, the interaction physics of the system will be described by a long-range Coulomb interaction rather than an on-site Hubbard interaction~\cite{tolsma2017orbital,tolsma2016quasiparticle}. Since fractional modes emerge in the strong-coupling limit, it is necessary for the electron density to be small. Electrostatic gating of the double-wire system is a possible route to satisfying both $k_A - 3k_F = \delta$ and that the system lives in the strong-coupling limit, $k_F \propto r_S^{-1} << 1$. Besides satisfying the momentum conservation condition for the scattering processes $\mathcal{O}^{\mathrm{SC}}_{3b}$ and $\mathcal{O}^{\mathrm{BS}}_{3b}$, one or the other must have a large bare coupling constant. In the event that it is not possible to identify a material with strong bare values of $g^{\mathrm{SC}}_{3b}$ and $g^{\mathrm{BS}}_{3b}$, it seems more important that $\kappa_{C(S)}$ is chosen such that $g^{\mathrm{BS}}_{3b}$ is assumed to be the larger one. In this case it is possible that $g^{\mathrm{BS}}_{3b}$ is significantly greater than $\Delta^*$, which is important because $\mathcal{O}^{\mathrm{BS}}_{3b}$ does not commute with $\mathcal{O}^{\mathrm{SC}}_{\mathrm{sing.}}$. And even if it starts from a small bare coupling, $g^{\mathrm{SC}}_{3b}$ can still flow to strong coupling faster than $\Delta^*$ for choices of $\kappa_{C(S)}$ such that $\kappa_C^{-1} + \kappa_S > \kappa_C^{-1} + 9 \kappa_S < 1$.

\section{Summary And Discussion}
\label{Sect:Four}
We have proposed a platform for MZMs and fractional zero modes in tunnel-coupled quantum wires where Rashba spin-orbit coupling is either small or completely absent.The latter scenario is always realized in systems with no inversion asymmetry, neither in the bulk nor in the structure. As a first step, in Sec.~\ref{Sect:One} we demonstrated that one-dimensional helical fermions can appear in a highly tunable setup which consists of two tunnel-coupled quantum wires with opposite sign Land\'e $g$ factors. We demonstrated that the cooperative effect of interwire spin-conserving tunneling and a small magnetic field along the direction of the wire leads to an effective low-energy Hamiltonian describing electrons whose spin orientation flips for opposite momentum states (i.e., helical fermions). 

Before commenting on the effect of proximity induced superconductivity or strong correlations, let's discuss the realistic values of parameters for our model. For definiteness we consider the quantitative values used to obtain Fig.~{\ref{fig:Two}} (c), which clearly demonstrates helical fermions in the lowest energy of the four bands. Assuming a modest value for the perpendicular magnetic field, $B_z \sim 10 \, {\rm T}$, we have the following expression for the interwire separation distance, $y_0 = \sqrt{|g^*| m^*} \, 7.26 \,{\rm nm}$. Here $m^*$ is the ratio of the effective mass to the bare electron mass, and $g^*$ is the ratio of the effective Land\'e factor to the bare value in either wire (for simplicity we assume they have equal and opposite effective $g$ factors). As mentioned in the introduction, the effective $g$ factor depends on many influences (strain, confinement, many-body interactions, atomic spin-orbit interactions), but can vary by almost two orders of magnitude from the bare value~\cite{hota1991theory, hota1993theory, qu2016quantized, bi2018spin, wojnar2012giant,yu2010giant,song2015first}. And the effective mass is similarly tunable over a large range by selecting particular materials: from $m^* \sim 0.01$ in InSb to $m^* \sim 10$ for SrTiO$_3$ based nanowires~\cite{irvin2013anomalous}. Assuming large mass and $g$ factor of $m^*=10$ and $|g^*| = 50$, respectively, we find that our two wires should be separated by $y_0 \sim 160$ nm, which is within the range commonly used in a variety of double-quantum-wire experiments~\cite{debray2001experimental,wegscheider2006magnetotransport,yamamoto2015band,auslaender2002tunneling}. In addition to the possibility to experimentally realize our platform with quantum wires, nanotubes, nanoribbons, and lithographically defined 1D channels on two-dimensional electron gases, we believe our model could also be realized using tunnel-coupled 1D chains of adatoms analogous to Refs.~\cite{klinovaja2013topological, nadj2014observation}. 

In Sec.~\ref{Sect:Two} we described how the presence of an $s$-wave superconducting pairing potential in our model's Hamiltonian leads to emergent MZMs. Zero-bias conductance peaks in recent experiments on proximity coupled quantum wires with strong Rashba spin-orbit interactions have provided strong evidence for Majorana zero modes, and our proposal opens the possibility for observing these non-Abelian quasiparticles in a wider class of Hamiltonians (i.e., those with inversion symmetry). One notable weakness of our model is the dependence on interwire superconducting pairing. This type of pairing is still possible, however, when the size of Cooper pairs in the superconductor is larger than the interwire separation distance~\cite{hofstetter2009cooper,das2012high}. Moreover, it has recently been demonstrated that interwire pairing can exceed intrawire pairing in the presence of strong interactions~\cite{thakurathi2018majorana}. Enhancing the magnitude and penetration depth of the proximity-inducing superconducting pairing-gap remains an ongoing engineering problem with wide interest. An adatom approach can also be beneficial to finally observing in an experimental setting the interaction-induced fractionalization of the Majorana zero mode. For example, if constructed using the tightly bound $d$- or $f$- atomic-orbitals with significantly larger electron-electron interaction energy scales compared to the semiconductor-based 1D materials from which most quantum wires are built. 
\begin{acknowledgments}
J.R.T. thanks Fan Zhang for helpful discussions and acknowledges support from the Swiss National Science Foundation.  
\end{acknowledgments}

\appendix
\begin{widetext}
\section{Details of Projection from $\mathcal{H}_{{\rm BdG}}$ to $	\mathcal{H}^{\mathrm{eff}}_{\rm BdG}$}
\label{Sect:AppendixA}
In this Appendix, we give details of the projection from the $8\times8$ full Hamiltonian $\mathcal{H}_{\rm BdG}$ to $4\times4$ effective low-energy description $\mathcal{H}^{\mathrm{eff}}_{\rm BdG}$ using a similar methodology as used for obtaining \eqref{eq:ProjectionWire}. First, we again start by rearranging the terms in  $\mathcal{H}_{\rm BdG}$ into $4\times4$ blocks given in Eq.~\eqref{eq:BlockTwo} where matrices  $\mathcal{\widetilde{H}}_{{\rm BdG},11}$  and $\mathcal{\widetilde{H}}_{{\rm BdG},22} $ corresponds to the bases $(\psi_{1\downarrow},\psi_{2\uparrow},\psi^\dagger_{1\downarrow},\psi^\dagger_{2\uparrow})^T$ and $(\psi_{1\uparrow},\psi_{2\downarrow},\psi^\dagger_{1\uparrow},\psi^\dagger_{2\downarrow})^T$ respectively. In this rearrangement the explicit form of $\mathcal{\widetilde{H}}_{{\rm BdG}}$ is
\small
\begin{align}
\renewcommand{\arraystretch}{1.1}
\mathcal{\widetilde{H}}_{{\rm BdG}}=\begin{bmatrix*}
\xi_- -B_z & 0 & 0 & -\Delta_2 &B_x & t& -\Delta_1 & 0 \\
0 & \xi_+-B_z & \Delta_2&0 &t & B_x & 0 & \Delta_1\\
0 & \Delta_2 &B_z -\xi_+ & 0  &\Delta_1 & 0 & -B_x &-t \\
-\Delta_2& 0 &0 & B_z -\xi_-&0& -\Delta_1 &-t & -B_x\\
B_x & t& -\Delta_1 & 0&\xi_- +B_z & 0 & 0 & \Delta_2  \\
t & B_x & 0 & \Delta_1&0 & \xi_++B_z & -\Delta_2&0 \\
-\Delta_1 & 0 & -B_x &-t&0 & -\Delta_2 &-B_z -\xi_+ & 0 \\
0& \Delta_1 &-t & -B_x&\Delta_2& 0 &0 & -B_z -\xi_-\\
\end{bmatrix*}.	\label{eq:Hrearr}
\end{align}
\normalsize
We then apply second order perturbation theory
\begin{align}
\mathcal{H}^{\mathrm{eff}}_{\rm BdG} = \mathcal{\widetilde{H}}_{{\rm BdG},11} \, + \, \mathcal{\widetilde{H}}_{{\rm BdG},12}\left[{\mathcal{\widetilde{H}}_{{\rm BdG},22}}\right]^{-1}\mathcal{\widetilde{H}}_{{\rm BdG},21},
\end{align}
which delivers the effective $4\times4$ Hamiltonian
\small
\begin{align}
\renewcommand{\arraystretch}{1.1}
\mathcal{H}^{\mathrm{eff}}_{\rm BdG}= \begin{bmatrix*}
\xi_- -B_z + E_1 & B_x C_1 & p_x C_2 & -\Delta_2 - C_3  \\
B_x C_1  & \xi_+-B_z+E_2& \Delta_2 + C'_3&p_xC'_2 \\
p_x C_2 & \Delta_2 + C'_3 &B_z -\xi_+ +E_3 & -BxC'_1  \\
-\Delta_2-C_3& p_xC'_2 &-BxC'_1 & B_z -\xi_-+ E_4\\\label{eq:Heff2}
\end{bmatrix*},
\end{align}
\normalsize
where $\xi_\pm = (p_x \pm A_x)^2 -\mu$, $E_i(i=1,2,3,4)$ are the corrections to the diagonal dispersions and $C_i, C'_i$ with $i=1,2,3$ are corrections to the coupling terms. The explicit equations for $E_i$, $C_i$ and $C'_i$ 
are far too complicated to be given here. We note that Eq. 
\eqref{eq:Heff2} is not quite what is given in Eq. \eqref{eq:LinearEff}. 
In particular, we have $E_i\neq E_j$ and $C_i \neq C'_j$for $i\neq j$. 
It turns out that difference $E_i-E_j$
is of higher order in $p_x$ and upon expanding around $p_x=0$ 
vanish. Linearizing at this $p_x=0$ delivers	
\begin{equation}
\begin{split}
\mathcal{H}^{\mathrm{eff}}_{\rm BdG}\left(p_x \sim 0 \right)=\left[ {\begin{array}{cccc}
-u^*p_x-\mu^* & t^* &\Lambda p_x & -\Delta'^* \\
t^* &	u^*p_x-\mu^* &  \Delta^* & \Lambda p_x \\
\Lambda p_x & \Delta^* & 	-u^*p_x+\mu^* &-t^* \\
-\Delta'^*& \Lambda p_x & -t^* &	u^*p_x+\mu^* \\
\end{array} } \right],\label{eq:LinHam2}
\end{split}
\end{equation}
where
\small
\begin{align}
&
\mu^* 
= \frac{2 m \left(A_x^2 \left(t^2-\Delta _1^2\right)+2 m \left(B_z \left(t^2-\Delta _1^2\right)+\Delta _1^2 \mu -\mu  t^2+2 \Delta _1 \Delta _2 t\right)\right)}{4 m A_x^2 \left(B_z-\mu \right)+A_x^4+4 m^2
\left(-2 \mu  B_z+B_z^2+\Delta _2^2+\mu ^2\right)}\label{eq:mu*}\nonumber\\
&
\phantom{\mu^*=}
+\frac{2 m B_x^2 \left(A_x^2+2 m \left(B_z-\mu \right)\right)}{4 m A_x^2 \left(B_z-\mu \right)+A_x^4+4 m^2 \left(-2 \mu  B_z+B_z^2+\Delta _2^2+\mu^2\right)}+\frac{A_x^2}{2 m}-B_z-\mu,
\end{align}
\begin{align}
&
u^* = \frac{4 m A_x \left(\Delta _1^2-t^2\right)}{4 m A_x^2 \left(B_z-\mu \right)+A_x^4+4 m^2 \left(-2 \mu  B_z+B_z^2+\Delta _2^2+\mu ^2\right)}\notag\\
&
\phantom{u^* = }
+\frac{8 m \left(2 m A_x \left(B_z-\mu \right)+A_x^3\right)
\left(A_x^2 \left(t^2-\Delta _1^2\right)+2 m \left(B_z \left(t^2-\Delta _1^2\right)+\Delta _1^2 \mu -\mu  t^2+2 \Delta _1 \Delta _2 t\right)\right)}{\left(4 m A_x^2 \left(B_z-\mu \right)+A_x^4+4 m^2
\left(-2 \mu  B_z+B_z^2+\Delta _2^2+\mu ^2\right)\right){}^2}\label{eq:u*}\notag\\
&
\phantom{u^* = }
-\frac{4 m A_x B_x^2 \left(4 m A_x^2 \left(B_z-\mu \right)+A_x^4+4 m^2 \left(-2 \mu  B_z+B_z^2-\Delta _2^2+\mu ^2\right)\right)}{\left(4 m
A_x^2 \left(B_z-\mu \right)+A_x^4+4 m^2 \left(-2 \mu  B_z+B_z^2+\Delta _2^2+\mu ^2\right)\right){}^2}+\frac{A_x}{m},
\end{align}
\begin{align}
&
t^* =\frac{4 m B_x \left(t A_x^2+2 m \left(t B_z+\Delta _1 \Delta _2-\mu t\right)\right)}{4 m A_x^2 \left(B_z-\mu \right)+A_x^4+4 m^2 \left(-2 \mu  B_z+B_z^2+\Delta _2^2+\mu ^2\right)},\label{eq:t*}
\end{align}
\begin{align}
&
\Lambda  =\frac{8 m A_x B_x \left(4 m A_x^2 \left(\Delta _1 B_z-\Delta _1\mu-\Delta _2 t\right)+\Delta _1 A_x^4\right)}{\left(4 m A_x^2 \left(B_z-\mu \right)+A_x^4+4 m^2 \left(-2 \mu  B_z+B_z^2+\Delta _2^2+\mu ^2\right)\right){}^2}\label{eq:Lambda}\notag\\
&
\phantom{\Lambda  =}
+\frac{8 m A_x B_x \left(4 m^2 \left(-2 B_z \left(\Delta _1 \mu +\Delta _2 t\right)+\Delta _1 B_z^2+\Delta _1 \mu
^2-\Delta _1 \Delta _2^2+2 \Delta _2 \mu  t\right)\right)}{\left(4 m A_x^2 \left(B_z-\mu \right)+A_x^4+4 m^2 \left(-2 \mu  B_z+B_z^2+\Delta _2^2+\mu ^2\right)\right){}^2},
\end{align}
\begin{align}
&
\Delta^* = \Delta_2
-\frac{4 m \left(m \left(-2 \Delta _1 t B_z-\Delta _1^2
\Delta _2+\Delta _2 t^2+2 \Delta _1 \mu  t\right)-\Delta _1 t A_x^2\right)}{4 m A_x^2 \left(B_z-\mu \right)+A_x^4+4 m^2 \left(-2 \mu  B_z+B_z^2+\Delta _2^2+\mu ^2\right)}
+\frac{4 \Delta _2 m^2B_x^2}{4 m A_x^2 \left(B_z-\mu \right)+A_x^4+4 m^2 \left(-2 \mu  B_z+B_z^2+\Delta _2^2+\mu ^2\right)}\notag\\
&
\phantom{\Delta^* =}
+ \frac{8 m A_x p_x \left(2 m A_x^2 \left(2 \Delta _1 t B_z-\Delta _2 B_x^2+\Delta _1^2 \Delta _2-\Delta _2 t^2-2 \Delta _1 \mu  t\right)+\Delta _1 t A_x^4\right)}{\left(4 m A_x^2 \left(B_z-\mu \right)+A_x^4+4 m^2 \left(-2 \mu  B_z+B_z^2+\Delta _2^2+\mu ^2\right)\right){}^2}\notag\\
&
\phantom{\Delta^* =}
+\frac{8 m A_x p_x \left(4 m^2 \left(-B_z \left(\Delta _2 B_x^2-\Delta
_1^2 \Delta _2+\Delta _2 t^2+2 \Delta _1 \mu  t\right)+\Delta _1 t B_z^2+\Delta _2 \mu  \left(B_x^2-\Delta _1^2\right)+\Delta _2 \mu  t^2+\Delta _1 t \left(\mu ^2-\Delta
_2^2\right)\right)\right)}{\left(4 m A_x^2 \left(B_z-\mu \right)+A_x^4+4 m^2 \left(-2 \mu  B_z+B_z^2+\Delta _2^2+\mu ^2\right)\right){}^2}\label{eq:delta}
\end{align}
\begin{align}
&
\Delta'^*= \Delta_2
-\frac{4 m \left(m \left(-2 \Delta _1 t B_z-\Delta _1^2
\Delta _2+\Delta _2 t^2+2 \Delta _1 \mu  t\right)-\Delta _1 t A_x^2\right)}{4 m A_x^2 \left(B_z-\mu \right)+A_x^4+4 m^2 \left(-2 \mu  B_z+B_z^2+\Delta _2^2+\mu ^2\right)}
+\frac{4 \Delta _2 m^2B_x^2}{4 m A_x^2 \left(B_z-\mu \right)+A_x^4+4 m^2 \left(-2 \mu  B_z+B_z^2+\Delta _2^2+\mu ^2\right)}\notag\\
&
\phantom{\Delta'^* =}
- \frac{8 m A_x p_x \left(2 m A_x^2 \left(2 \Delta _1 t B_z-\Delta _2 B_x^2+\Delta _1^2 \Delta _2-\Delta _2 t^2-2 \Delta _1 \mu  t\right)+\Delta _1 t A_x^4\right)}{\left(4 m A_x^2 \left(B_z-\mu \right)+A_x^4+4 m^2 \left(-2 \mu  B_z+B_z^2+\Delta _2^2+\mu ^2\right)\right){}^2}\notag\\
&
\phantom{\Delta'^* =}
-\frac{8 m A_x p_x \left(4 m^2 \left(-B_z \left(\Delta _2 B_x^2-\Delta
_1^2 \Delta _2+\Delta _2 t^2+2 \Delta _1 \mu  t\right)+\Delta _1 t B_z^2+\Delta _2 \mu  \left(B_x^2-\Delta _1^2\right)+\Delta _2 \mu  t^2+\Delta _1 t \left(\mu ^2-\Delta
_2^2\right)\right)\right)}{\left(4 m A_x^2 \left(B_z-\mu \right)+A_x^4+4 m^2 \left(-2 \mu  B_z+B_z^2+\Delta _2^2+\mu ^2\right)\right){}^2}.\label{eq:deltap}
\end{align}	
\normalsize
We observe that linearized effective Hamiltonian $\mathcal{H}^{\mathrm{eff}}_{\rm BdG}$ has almost the same form as Hamiltonian Eq. \eqref{eq:LinearEff} except the modified pairing potentials $\Delta^*$ and $\Delta'^*$. Focusing on these terms,
we realize that the only difference between them is the last two terms having opposite signs. Note that both of these terms also have a $p_x$ dependence which is ignored in the Hamiltonian Eq. \eqref{eq:LinearEff}.
We justify omitting these terms, demanding that $B_z$ is the largest energy scale in the model. Therefore, the denominator of the difference $\Delta^*-\Delta'^*$
is negligible compared to the other terms. Therefore, we take $\Delta^* \approx \Delta'^*$ up to corrections $\mathcal{O}(p_x)$.

\section{Linearized Spectrum}
\label{Sect:AppendixB}

\begin{figure}[H]
\includegraphics[width=0.95\linewidth]{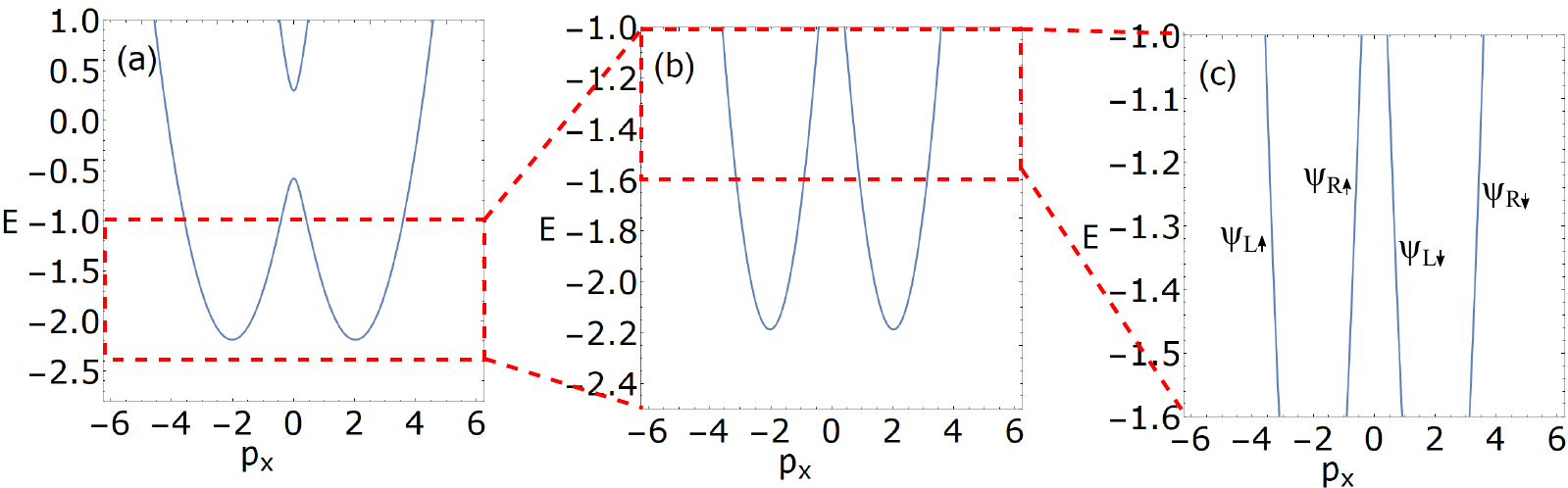}
\caption{(Color online) Evolution of the lowest lying band given in Fig.~\ref{fig:Three}(b) as we zoom into smaller energy ranges. (a) Choosing an appropriate chemical potential allows us to stay away from the gap. (b) Focusing on the energies below the gap leads to two disjoint bands where there are four Fermi points at $\pm4k_F,\pm2k_F$. (c) We linearize at these four Fermi points to obtain two right- and two left-movers of opposite spin.}\label{fig:Six}
\end{figure}

Finally, we address how the lowest energy bands is linearized around four Fermi points. For demonstrative purposes we start with the band structure in Fig.~\ref{fig:Three}(b). In Fig.~\ref{fig:Six} (a) the spectrum zoomed around the lowest energy band is shown where the second lowest-energy level is also present. Choosing an appropriate chemical potential and focusing on energies below the gap, say $E~(-2.5,-1)$ as in Fig.~\ref{fig:Six}(b), we then observe that the spectrum reduces effectively to two bands that are populated by helical fermions, see Section \ref{Sect:One}. We can linearize at four Fermi points given by $\pm k_A \pm k_F = \pm4k_F,\pm2k_F$, which results in two right- and two left-mover modes of opposite spin as shown in Fig.~\ref{fig:Six} (c). 
Finally, these fermionic modes can be bosonized to yield fractional modes
upon adding interactions.

\end{widetext}
\bibliography{References}
\end{document}